\documentclass{article}

\pdfoutput=1

\usepackage{arxiv}

\usepackage[utf8]{inputenc}
\usepackage[T1]{fontenc}
\usepackage[numbers,sort&compress]{natbib}
\usepackage{hyperref}
\usepackage{url}
\usepackage{booktabs}
\usepackage{amsfonts}
\usepackage{nicefrac}
\usepackage{microtype}
\usepackage{graphicx}
\usepackage{doi}
\usepackage{array}
\usepackage{xcolor}
\usepackage{multirow}
\usepackage{makecell}
\usepackage{pdfcomment}   
\usepackage{float}        
\usepackage{needspace}    

\title{Why AI Readiness Is an Organizational Learning Problem,\\Not a Technology Purchase}

\author{%
  Jeanne McClure, PhD\\
  Ars Innovate Technology and Consulting\\
  NC State University\\
  \texttt{jmcclure@arsinnovate.com}
  \And
  Gregg Gerdau\\
  Matador Advisors\\
  \texttt{gregg@matadoradvisors.com}
}

\date{2026}

\renewcommand{\headeright}{McClure \& Gerdau, 2026}
\renewcommand{\undertitle}{The Structural Evolution of Enterprise AI Capability}
\renewcommand{\shorttitle}{\textit{The Structural Evolution of Enterprise AI Capability}}

\hypersetup{
  pdftitle={Why AI Readiness Is an Organizational Learning Problem, Not a Technology Purchase},
  pdfsubject={Enterprise AI, Organizational Readiness, AI Strategy},
  pdfauthor={Jeanne McClure, Gregg Gerdau},
  pdfkeywords={AI readiness, enterprise AI, organizational learning, AI maturity, technology strategy},
}

\begin{document}
\raggedbottom

\maketitle

\begin{abstract}
Global corporate AI investment reached \$252.3 billion in 2024, yet only 6\% of firms report significant earnings impact. This article argues that AI project failure is fundamentally an organizational learning problem rather than a technology deficit. Drawing on a systematic synthesis of 20 foundational sources, including surveys of nearly 10,000 organizational leaders, we identify critical failure vectors across organizational culture, leadership alignment, and technical infrastructure. We introduce the Orchestration Maturity Framework, a model that maps enterprise AI capability across five pillars---Culture \& Leadership, Human Capital \& Operations, Data Architecture, Systems \& Infrastructure, and Governance \& Regulatory Compliance. The framework provides prescriptive guidance for transitioning from isolated experimentation to an orchestrated agentic enterprise, reframing AI investment as a multi-dimensional capability development challenge rather than a static IT procurement.
\end{abstract}


\section{The \$252 Billion Paradox}

Global corporate investment in artificial intelligence reached \$252.3 billion in 2024, reflecting a 13-fold expansion since 2014 \citep{maslej2025ai}. The divergence between capital allocation and realized value is not merely a technical lag; it is the visible result of specific, diagnosable organizational and functional frictions (see Table~\ref{tab:failure_domains}). While total corporate investment reached an all-time infrastructure peak of \$360.7B in 2021, the 2024 resurgence to \$252.3B represents a second wave of capital that is meeting the same structural resistance \citep{maslej2025ai, ransbotham2025emerging}. Yet McKinsey's 2025 global survey of nearly 2,000 respondents reveals that only 39\% of organizations report any enterprise-level earnings before interest and taxes (EBIT) impact from AI. Furthermore, just 6\% qualify as high performers seeing significant value, which the study defines as those attributing more than 20\% of their EBIT specifically to AI integration \citep{singla2025state}. According to Cisco's AI Readiness Index, half of surveyed organizations with more than 500 employees now allocate between 10\% and 30\% of their total IT budgets to AI, and 92\% of companies plan to increase their AI investments over the next three years \citep{cisco2024readiness, mayer2025superagency}. By most available measures, AI has moved from a discretionary experiment to a central line item in enterprise technology strategy.

\begin{figure}[H]
    \centering
    {\large\textbf{The \$252B Paradox}}\\[2pt]
    {\textbf{Capital Expansion vs.\ Organizational Resistance}}\\[6pt]
    \pdftooltip{%
      \includegraphics[width=1\linewidth]{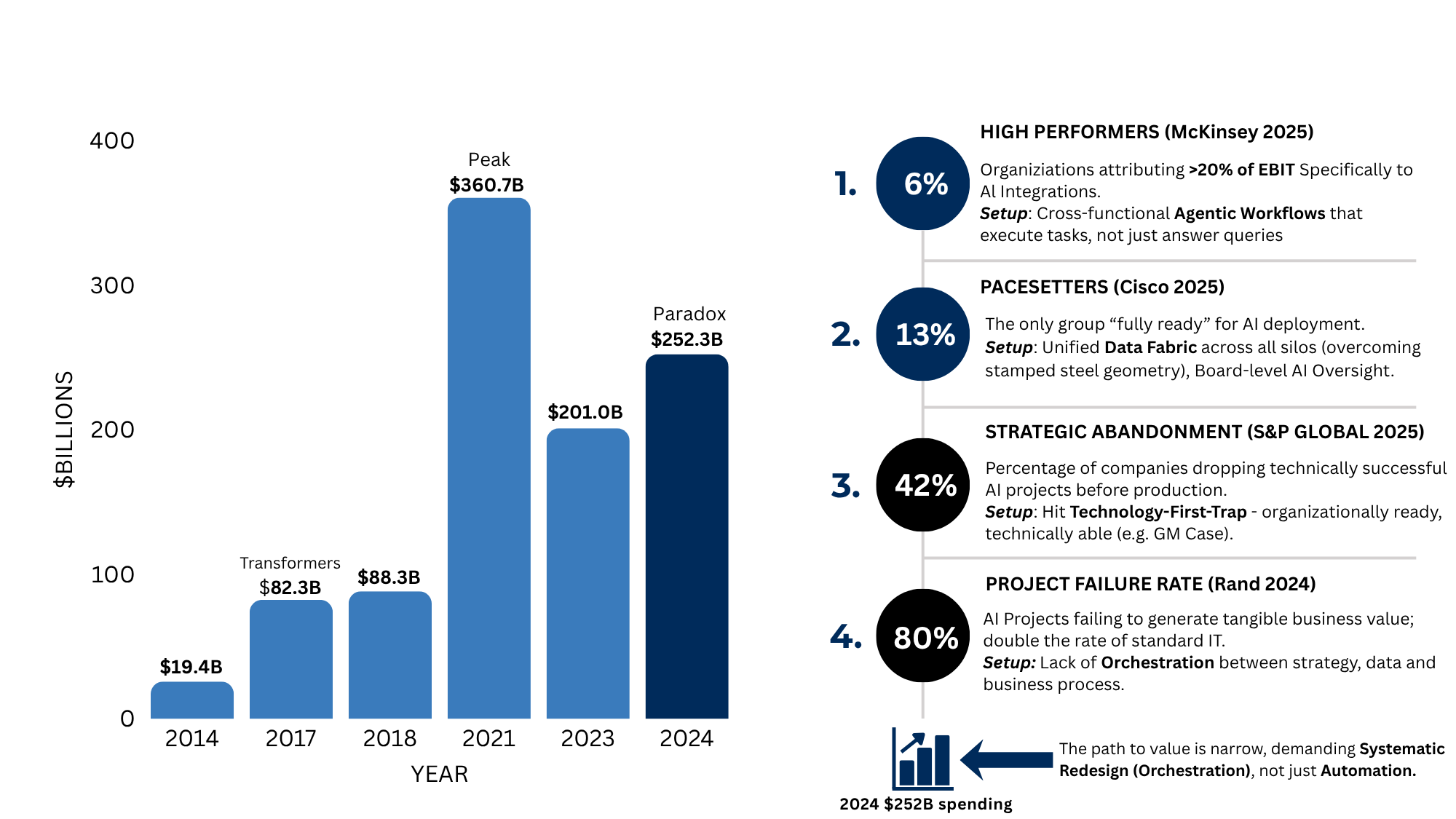}%
    }{Figure 1. Bar chart titled The \$252B Paradox: Capital Expansion vs. Organizational Resistance. The x-axis shows years 2014 to 2024. The y-axis shows billions of dollars. Bars rise from \$19.4B in 2014 to a peak of \$360.7B in 2021, then decline to \$252.3B in 2024. Annotated callouts highlight four data points: 6\% of firms are high performers attributing more than 20\% of EBIT to AI; 13\% are Pacesetters fully ready for deployment; 42\% of companies abandoned technically successful AI projects before production; and 80\% of AI projects fail to generate tangible business value, double the rate of standard IT initiatives.}
    \caption{Global AI investment versus organizational value realization. While
    capital allocation reached \$252.3B in 2024, only 6\% of firms achieved significant EBIT
    impact. The 80\% failure rate underscores a Technology First Trap where infrastructure
    and orchestration, not funding, remain the primary constraints.\\
    \textit{Sources: McKinsey (2025); Cisco AI Readiness (2025); S \& P Global (2025); Rand Corp (2024)}}
    \label{fig:paradox}
\end{figure}

This investment-value gap is quantified in Figure~\ref{fig:paradox}, illustrating the concentration of capital against a fragmented landscape of organizational readiness.

Where does this investment go? McKinsey data reveal a telling pattern: organizations report the greatest cost benefits from AI in software engineering (56\% reporting cost decreases), manufacturing (56\%), and IT (54\%), functions where AI automates well-defined tasks. Revenue benefits concentrate in marketing and sales (67\%), strategy and corporate finance (65\%), and product development (62\%), functions requiring cross-functional coordination \citep{singla2025state}. This divergence underscores the organizational challenge: capturing AI's full value requires not just automation of individual tasks but redesign of how functions work together.

As illustrated in Figure~\ref{fig:paradox}, the returns have not kept pace with spending. Estimates suggest that over 80\% of AI projects fail, roughly double the rate of non-AI IT initiatives \citep{ryseff2024root, ali2025factors}. S\&P Global's 451 Research reports that the share of companies abandoning the majority of their AI initiatives before production rose from 17\% to 42\% year over year \citep{johnston2025generative}. BCG finds that only 26\% of companies have moved beyond proof of concept to generate tangible value \citep{debellefonds2024value}. McKinsey reports that just 1\% of C-suite leaders describe their generative AI deployments as mature \citep{mayer2025superagency}.

The pattern is not one of insufficient investment. Organizations are spending more on AI than at any point in the technology's history, yet according to the reports readiness scores are declining, project abandonment is rising, and nearly half of business leaders report no gains or results below expectations \citep{cisco2024readiness, mayer2025superagency}. Israeli and Ascarza describe this in Harvard Business Review as a ``technology-first trap'': organizations deploy AI solutions department by department without linking them to enterprise goals, producing technically successful implementations that never reach production \citep{israeli2025most}.

The specific technical and cultural frictions contributing to these failure rates are categorized in Table~\ref{tab:failure_domains}, providing a diagnostic lens for the ``Technology-First Trap'' illustrated by the following case:

\begin{quote}
General Motors applied generative-design software to produce a seat bracket 40\% lighter and 20\% stronger than the original, yet the part never entered production because GM's supply chain, built for stamped steel, could not accommodate the AI-generated geometry. The technology worked. The organization was not ready.
\end{quote}

If funding is not the binding constraint, what is?

\needspace{6\baselineskip}
\begin{center}
  {\large\textbf{The Technology-First Trap}}\\[3pt]
  \textbf{Diagnostic Markers of AI implementation Failure}
\end{center}

\begin{table}[H]
\centering
\small
\begin{tabular}{p{3.0cm} p{3.0cm} p{4.8cm} p{3.8cm}}
\toprule
\textbf{PRIMARY FAILURE DOMAIN} &
\textbf{SPECIFIC DIAGNOSTIC INDICATOR} &
\textbf{EVIDENCE/IMPACT} &
\textbf{SUPPORTING SOURCES} \\
\midrule

\textbf{ORGANIZATIONAL}\newline(Structure) &
Leadership \& Strategic Alignment &
84\% of practitioners cite leadership-driven issues (objectives/ROI) as the root cause of project failure.\textsuperscript{6, 12} &
Ryseff et al.\ (RAND); Hoque, Davenport, \& Nelson (MIT SMR) \\
\addlinespace

\textbf{ORGANIZATIONAL}\newline(Structure) &
Cultural Resistance \& Change &
91\% of data leaders identify culture as the primary impediment; only 9\% point to technology.\textsuperscript{8, 12} &
Johnston (S\&P Global); Hoque, Davenport, \& Nelson (MIT SMR) \\
\addlinespace

\textbf{ORGANIZATIONAL}\newline(Structure) &
Governance \& Policy Regression &
Readiness fell from 44\% to 42\% as investment grew, revealing a lag in formal mitigation.\textsuperscript{4, 15} &
Cisco; Wade et al.\ (MIT SMR) \\
\addlinespace

\textbf{ORGANIZATIONAL}\newline(Structure) &
Human-AI Learning Deficit &
Learning increases benefit likelihood by 34\%, nearly 2x more effective than infrastructure (19\%).\textsuperscript{2, 19} &
Ransbotham et al.\ (MIT SMR/BCG); Pinski et al.\ (CHI) \\
\addlinespace

\textbf{ORGANIZATIONAL}\newline(Structure) &
The Scaling Paradox &
Project abandonment spiked from 17\% to 42\% as organizations hit structural entanglement.\textsuperscript{8, 10, 11} &
\r{A}str\"{o}m et al.\ (CMR); Johnston (S\&P Global); Israeli \& Ascarza (HBR) \\
\addlinespace

\textbf{TECHNICAL}\newline(Functional) &
The Semantic Bottleneck &
90\% of effort is lost to manual data prep due to a lack of automated ontologies and metadata.\textsuperscript{9, 14, 16} &
de Bellefonds et al.\ (BCG); Sadiq et al.\ (PeerJ); Modern Data Co. \\
\addlinespace

\textbf{TECHNICAL}\newline(Functional) &
The Output Avalanche &
Frictionless infrastructure can intensify workloads when output volume outpaces human review capacity.\textsuperscript{1, 20} &
Maslej et al.\ (Stanford HAI); Ranganathan \& Ye (HBR) \\

\bottomrule
\end{tabular}
\medskip
\caption{Synthesized Organizational vs Technical Causes of AI Project Failure.
A diagnostic breakdown of the Technology-First Trap, illustrating how 91\% of
implementation challenges are rooted in cultural leadership misalignment rather than
technical limitations.}
\label{tab:failure_domains}
\end{table}

\section{The Misdiagnosis}

RAND's study of AI project failure identified five root causes through interviews with 65 data scientists and engineers: leadership-driven failures, data quality limitations, bottom-up technology chasing, underinvestment in deployment infrastructure, and attempts to apply AI to problems beyond the current state of the art \citep{ryseff2024root}. Of these five, only the last reflects a genuinely technical limitation. Within that sample, 84\% of industry interviewees cited leadership-driven issues as the primary cause of failure \citep{ryseff2024root}.

This is not an artifact of a single study. In a global survey of 2,525 decision-makers, 91\% encountered implementation challenges across technological, organizational, and cultural domains simultaneously \citep{angstrom2023getting}. Hoque, Davenport, and Nelson report in MIT Sloan Management Review that 91\% of large-company data leaders identified cultural and change management challenges as the primary impediment to becoming data-driven; only 9\% pointed to technology \citep{hoque2025ai}. S\&P Global found that organizations with above-average failure rates were significantly more likely to report failures of communication and targeting rather than failures of the technology itself \citep{johnston2025generative}.

If the root causes are organizational, readiness assessments should show improvement as organizations invest. They do not. Cisco's AI Readiness Index, surveying 7,985 senior leaders, found that governance readiness fell year over year (from 44\% to 42\%), cultural readiness dropped sharply (from 40\% to 31\% among Chasers), and board receptivity to AI fell from 82\% to 66\%, all while AI budgets grew \citep{cisco2024readiness}. Experience does not simplify the challenge; it reveals deeper layers.

AI-experienced firms reported more challenges related to technical fit and regulatory compliance than newcomers, not fewer \citep{angstrom2023getting}. As one study observed, distinct technological challenges became ``organizationally entangled'' as firms matured \citep{angstrom2023getting}. Among organizations with extensive agentic AI adoption, 66\% anticipate fundamental changes to their operating models, compared to just 42\% with no adoption plans \citep{ransbotham2025emerging}.

\section{Why the C-Suite Cannot Delegate This}

BCG's global survey found that 62\% of realized AI value derives from core business functions such as operations, sales, and R\&D, while IT departments generate just 7\% \citep{debellefonds2024value}. The organizations generating the most value allocate resources accordingly: 70\% to people and processes, 20\% to technology, and just 10\% to algorithms \citep{debellefonds2024value, woerner2025grow}. Yet many organizations continue to treat AI primarily as an efficiency tool housed within IT.

AI has no single owner, and that is the problem. MIT CISR's research finds that the transition from pilots to scaled adoption requires joint leadership from the CEO, CIO, chief strategy officer, and head of human resources working as a coordinated team \citep{woerner2025grow, debellefonds2024value}. Without that coalition, organizations remain stuck in the pilot stage \citep{woerner2025grow, debellefonds2024value, johnston2025generative}.

The failure pattern points to the seam between functions: RAND's practitioners identified misalignment between business leaders who define the problem and technical teams who build the solution as the primary root cause \citep{ryseff2024root, mayer2025superagency}.

Primary responsibility for AI oversight is dispersed across information security (21\%), data and analytics (17\%), and dedicated AI governance roles (14\%), with no single function holding a dominant mandate \citep{maslej2025ai}. The governance infrastructure to manage these cross-functional demands is not yet in place.

Only 31\% of organizations report comprehensive AI policies, and AI policy ranked last among 23 readiness factors identified across 52 studies \citep{cisco2024readiness, ali2025factors}. Data fragmentation compounds the problem: 82\% of organizations report that their data remains siloed, while 88\% struggle with AI development disconnected from business operations \citep{cisco2024readiness, nemko2025maturity, angstrom2023getting}.

The organizations moving beyond pilots are restructuring leadership. At Italgas, Europe's largest natural gas distributor, the chief human resources officer role was redefined as the chief people, innovation, and transformation officer, with a new AI director reporting jointly to this role and the CIO \citep{woerner2025grow}. Guardian Life Insurance reorganized technology teams into cross-functional units with end-to-end accountability by product \citep{woerner2025grow}. JPMorgan Chase expanded its global head of data and analytics role to connect AI implementation to organizational purpose and ethics; PepsiCo and Standard Chartered created enterprise-wide transformation officer positions \citep{hoque2025ai}. These organizations generate measurably stronger financial performance: 50\% higher revenue growth and 40\% higher return on invested capital relative to industry peers \citep{debellefonds2024value, woerner2025grow}. McKinsey finds that organizations reporting the highest EBIT impact are 2.8 times more likely to have fundamentally redesigned their workflows \citep{singla2025state}.

Wade et al.\ warn that a dedicated C-suite AI role can create as many problems as it solves: in one case, a CDO office expanded to more than 100 full-time equivalents, turning all business case calculations negative \citep{wade2024chief}. Their conclusion is that the CAIO role, when warranted, should be a fixed-term appointment with a specific brief to build capabilities that are ultimately handed over to existing organizations \citep{wade2024chief}. The need for strategic altitude does not require a permanent C-suite addition. It requires the right mechanism for the organization's current stage.

\section{Readiness Is Not a Score. It Is a Progression.}

A growing number of frameworks assess organizational AI capabilities, but Sadiq et al.'s review found that 13 of 15 are descriptive, designed to assess the status quo, while only six offer prescriptive guidance for advancing between stages \citep{sadiq2021maturity}. They can tell an organization it is at level two but cannot explain what it would need to learn, build, or change to reach level three.

\begin{figure}[H]
    \centering
    {\textbf{Siloed-Integrated-Orchestrated Progression Across Five Pillars}}\\[6pt]
    \pdftooltip{%
      \includegraphics[width=1\linewidth]{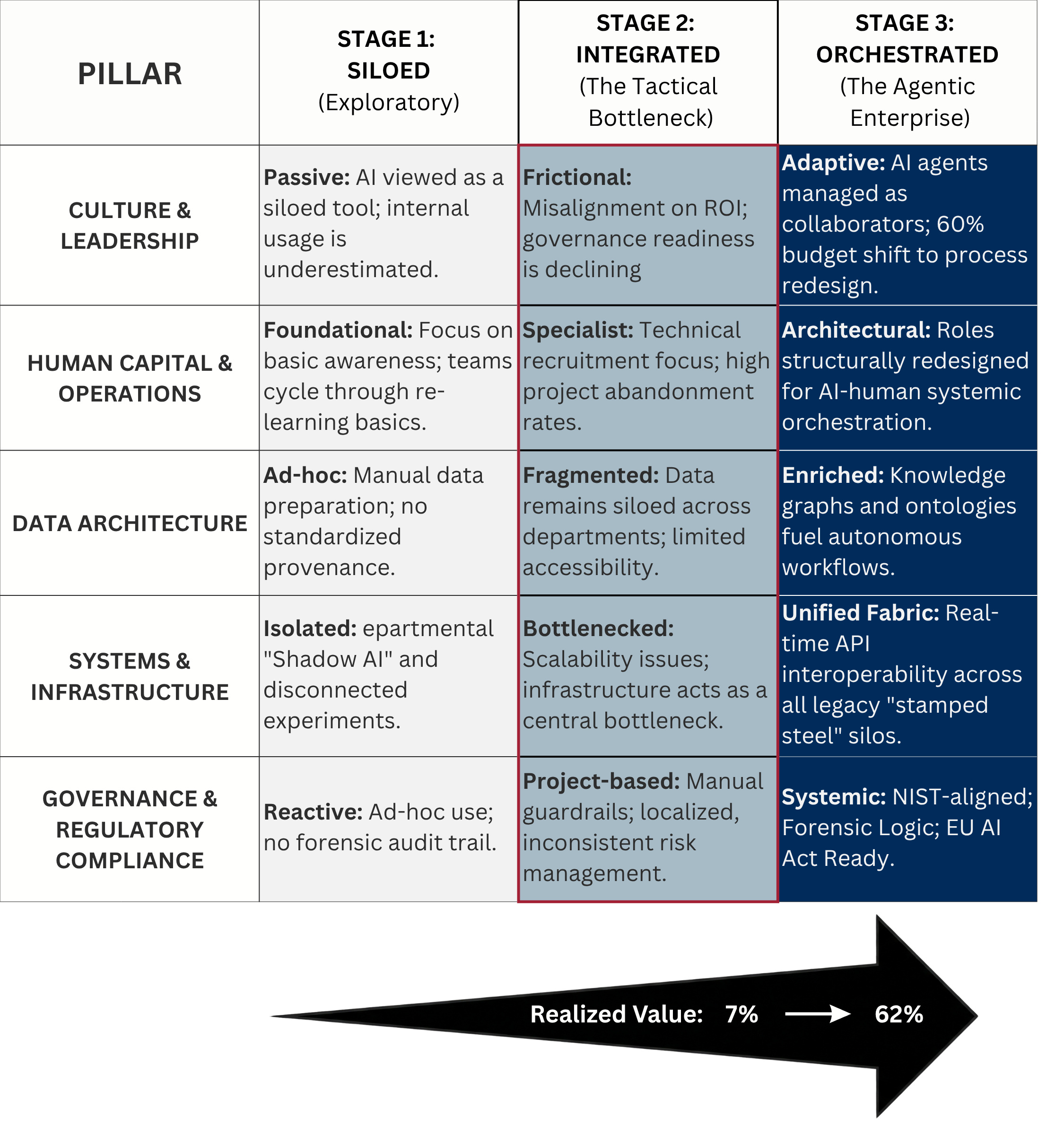}%
    }{Figure 2. Heatmap table titled Siloed-Integrated-Orchestrated Progression Across Five Pillars. Rows represent five pillars: Culture and Leadership, Human Capital and Operations, Data Architecture, Systems and Infrastructure, and Governance and Regulatory Compliance. Columns represent three stages: Stage 1 Siloed (Exploratory), Stage 2 Integrated (The Tactical Bottleneck), and Stage 3 Orchestrated (The Agentic Enterprise). Each cell describes the characteristic state of that pillar at that stage. An arrow at the bottom indicates realized value increases from 7\% at the Siloed stage to 62\% at the Orchestrated stage.}
    \caption{The Structural Evolution of Enterprise AI Capability. While global
    investment has expanded 13-fold since 2014, value realization remains a ``paradox''
    because most organizations treat AI as a static IT procurement rather than a
    multi-dimensional developmental challenge. This heatmap illustrates the shift from
    isolated, Siloed experimentation to an Orchestrated agentic enterprise.
    High-performing organizations are 2.8 times more likely to have fundamentally
    redesigned their workflows and governance structures to reach the Orchestrated stage.\\
    \textit{Source: Synthesized from NIST AI RMF, Cisco (2025), McKinsey (2025), and MITRE (2022).}}
    \label{fig:progression}
\end{figure}

The evidence suggests three qualitatively distinct phases which we term, Siloed, Integrated and Orchestrated (see Figure~\ref{fig:progression}), each defined by the nature of the architectural and cultural hurdles the organization faces.

The pillars are not new. Combining them with a developmental progression is.

At the \textbf{Siloed} stage, organizations are building basic understanding of what AI can and cannot do. Efforts are isolated, each initiative re-learns foundational skills, and leaders instruct technical teams to apply AI to problems that do not require it \citep{bloedorn2022mitre, ryseff2024root}. C-suite leaders estimate only 4\% of employees use generative AI extensively; actual usage runs roughly three times higher, often through unsanctioned tools \citep{mayer2025superagency, johnston2025generative}. These are not infrastructure problems. They are structural visibility problems.

At the \textbf{Integrated} stage, organizations can run pilots but many stall. Only 26\% have moved beyond proof of concept \citep{debellefonds2024value}, and 42\% abandoned the majority of their AI initiatives before production \citep{johnston2025generative}. This is the ``IT Silo Bottleneck.'' The transition to the next stage depends on leadership alignment: MIT CISR finds it requires a united top leadership team working jointly on strategy and synchronization to break through departmental friction \citep{woerner2025grow}.

At the \textbf{Orchestrated} stage, AI is embedded in how work gets done across the enterprise. The MIT SMR-BCG AI survey suggests this means viewing AI systems as coworkers requiring onboarding, performance oversight, and cross-functional governance \citep{ransbotham2025emerging}. Two-thirds of organizations with extensive AI adoption expect fundamental changes to their operating models \citep{ransbotham2025emerging}. Moderna has merged its technology and human resources departments to signal that AI agents must be managed as part of the workforce, not just IT infrastructure \citep{ransbotham2025emerging}. Orchestration requires institutional capabilities that cannot be purchased or installed but must be developed through sustained practice.

To navigate the structural progression illustrated in Figure~\ref{fig:progression}, we propose a model of structural evolution across five interconnected pillars of enterprise AI capability shifting the focus from transactional technology procurement to enduring capability: Culture \& Leadership \citep{ryseff2024root, hoque2025ai}; Human Capital \& Operations \citep{debellefonds2024value, woerner2025grow}; Data Architecture \citep{cisco2024readiness, moderndataco2025ai, nemko2025maturity}; Systems Infrastructure \citep{ryseff2024root, angstrom2023getting}; and Governance \& Regulatory Compliance \citep{cisco2024readiness, maslej2025ai}. While many organizations prioritize technical infrastructure, our research indicates that structural integrity across all five domains is required to move from isolated experiments to an Orchestrated agentic enterprise.

Within each stage, these five interconnected pillars present qualitatively different challenges: At the Siloed stage, data practices are ad hoc and manual preparation consumes up to 90\% of project effort \citep{moderndataco2025ai}; at the Integrated stage, fragmentation across boundaries is the bottleneck; at the Orchestrated stage, agentic AI systems require data enriched with ontologies, knowledge graphs, and contextual metadata \citep{angstrom2023getting, singla2025state}.

Infrastructure shifts from shadow AI to scalability bottleneck to orchestrating dynamic agentic workflows \citep{johnston2025generative, cisco2024readiness, ransbotham2025emerging}. Talent gaps evolve from foundational awareness to augmented domain expertise to workforce reimagination \citep{ryseff2024root, pinski2023ai, ransbotham2025emerging}. Culture shifts from employee fear to functional silos to flattened hierarchies and decentralized decision-making \citep{angstrom2023getting, mayer2025superagency, woerner2025grow}.

Finally, governance matures from reactive, ad-hoc usage characterized by a lack of formal mitigation and ``shadow AI'' adoption \citep{maslej2025ai, cisco2024readiness} to project-based manual guardrails where oversight is fragmented and readiness often regresses due to regulatory complexity \citep{wade2024chief, cisco2024readiness}. In the Orchestrated stage, governance becomes systemic and NIST-aligned, utilizing forensic logic and automated compliance to manage the unique risks of autonomous agentic workflows \citep{bloedorn2022mitre, ransbotham2025emerging, nemko2025maturity}.

\section{Stop Buying. Start Building.}

These implications follow from this evidence:

\medskip
\noindent\textbf{Reframe the problem before investing in solutions.} The data consistently show that AI failure is an organizational coordination problem that manifests as technology failure. Before launching the next AI initiative, leaders should ask whether the barriers they face involve cross-functional alignment, misaligned incentives, or capability gaps across roles. If they do, additional technology investment alone will not resolve them \citep{israeli2025most}.

\medskip
\noindent\textbf{Assess where you are on the progression, not just what you have.} Static readiness checklists measure whether an organization possesses certain capabilities. They do not measure whether the organization can learn, adapt, and improve its use of AI over time. An organization at the Siloed stage struggling with shadow AI faces a fundamentally different problem than one at the Integrated stage struggling to scale pilots. Organizations that remain at the Siloed stage while competitors progress face compounding costs: not just financial, but through organizational cynicism that undermines future initiatives and talent drain as skilled employees leave.

\medskip
\noindent\textbf{Ensure the enablement capability exists at the right altitude.} If 62\% of AI value originates in business functions rather than IT, and if 84\% of failures trace to leadership-driven causes, the capability to drive AI readiness must have cross-functional authority. Whether the solution takes the form of a dedicated chief AI officer, an expanded transformation mandate, or a cross-functional governance team depends on context. But it cannot live inside a single department \citep{debellefonds2024value, woerner2025grow, singla2025state}.

\medskip
\noindent\textbf{Recognize that the cost of inaction accelerates.} Ransbotham et al.'s ninth annual MIT SMR-BCG survey finds that traditional AI adoption has climbed to 72\%, agentic AI has already reached 35\%, and 76\% of respondents view agentic AI as more like a coworker than a tool \citep{ransbotham2025emerging}. Organizations that have not built the foundations for basic AI adoption now face a compounding challenge: they must simultaneously develop siloed-stage capabilities while competitors navigate the Orchestrated stage demands of agentic systems. The window for building foundational readiness is narrowing.

\medskip
\noindent\textbf{Match measurement to maturity.} Only 39\% of organizations report any enterprise-level EBIT impact from AI, and most attribute less than 5\% of EBIT to AI use \citep{singla2025state}. Yet AI high performers are 3.6 times more likely to pursue transformative change and 2.8 times more likely to have fundamentally redesigned their workflows \citep{singla2025state}. Organizations that measure only cost savings will optimize for incremental gains. Measurement frameworks should evolve with organizational maturity: from input metrics (training hours, tools deployed) at the Siloed stage, to process metrics (cross-functional coordination, time-to-deployment) at Integrated, to outcome metrics (AI-attributable revenue, workflow redesign impact) at Orchestrated.

\medskip
The \$252 billion paradox is not a technology problem awaiting a technology solution. It is an organizational learning problem that requires developmental, not transactional, intervention. The question facing leaders is no longer whether to invest in AI. It is whether their organizations can learn fast enough to make those investments count.

\section*{Acknowledgment}

This material is based upon work in part supported by the National Science Foundation under Grant \#DGE-2222148. Any opinions, findings, and conclusions or recommendations expressed in this material are those of the author(s) and do not necessarily reflect the views of the National Science Foundation.

\section*{Methodology and Authors' Note}

This synthesis draws on systematic literature review of 20 foundational sources, including large-scale industry surveys (Cisco, $n = 7{,}985$; McKinsey global survey, $n = 1{,}833$; the Stanford 2025 AI Index Report, and practitioner studies from the RAND Corporation). The framework is further informed by global standards including the NIST AI Risk Management Framework 1.0 and the MITRE AI Maturity Model.

\textit{Author note:} The developmental framing of the readiness model was informed by the authors' applied research across educational and enterprise settings, including work on structured human-AI collaboration and organizational AI adoption.


\end{document}